\documentclass[nature,twocolumn,superscriptaddress,showpacs,nofootinbib]{revtex4-1}

\usepackage{graphicx} 
\usepackage{amsmath} 
\usepackage{epstopdf}
\usepackage[abs]{overpic} 
\usepackage{subfigure} 
\usepackage{cancel}
\usepackage{nicefrac}

\begin{document}

	\title{The $(e_g \otimes e_u) \otimes E_g$ product Jahn-Teller effect in the neutral group-IV--vacancy quantum bits in diamond}
	
	\author{Gerg\H{o} Thiering} \affiliation{Wigner Research Centre for Physics,
		Hungarian Academy of Sciences, PO Box 49, H-1525, Budapest, Hungary}
	\affiliation{Department of Atomic Physics, Budapest University of Technology and
		Economics, Budafoki \'ut 8., H-1111 Budapest, Hungary}
	
	\author{Adam Gali} \email{gali.adam@wigner.mta.hu} \affiliation{Wigner Research
		Centre for Physics, Hungarian Academy of Sciences, PO Box 49, H-1525, Budapest,
		Hungary} \affiliation{Department of Atomic Physics, Budapest University of
		Technology and Economics, Budafoki \'ut 8., H-1111 Budapest, Hungary}

	\begin{abstract}
The product Jahn-Teller (pJT) effect may occur for such coupled electron-phonon systems in solids where single electrons occupy double degenerate orbitals. We propose that the excited state of the neutral $X$V split-vacancy complex in diamond, where $X$ and V labels a group-IV impurity atom of $X$=Si, Ge, Sn, Pb and the vacancy, respectively, is such a system with $e_g$ and $e_u$ double degenerate orbitals and $E_g$ quasi-localized phonons. We develop and apply \emph{ab initio} theory to quantify the strength of electron-phonon coupling for neutral $X$V complexes in diamond, and find a significant impact on the corresponding optical properties of these centers. Our results show good agreement with recent experimental data on the prospective SiV($0$) quantum bit, and reveals the complex nature of the excited states of neutral $X$V color centers in diamond. 
	\end{abstract}
	\maketitle

\section*{Introduction\label{sec:Intro}}
Fluorescent, paramagnetic point defects in diamond may realize quantum bits for quantum technology. Split-vacancy complexes of group-IV impurity atom ($X$=Si, Ge, Sn, Pb) and vacancy, i.e., $X$V defects with $D_{3d}$ symmetry are in the focus of intense research. The negatively charged $X$V, i.e., $X$V($-$) defects have $S=1/2$ spin state and fluoresce mostly in the visible~\cite{elso_jel_Zaitsev, Goss1996, Clark_SiV_1.68_1995, neu2011, Gali:PRB2013, iwasaki2015germanium, Ralchenko2015, Huler2017, tchernij2017single, iwasaki2017tin, Trusheim2018}. The inversion symmetry of the centers assumes virtually no Stark-shift in the optical signals which is a prerequisite for realization of indistinguishable single photon sources. Among these color centers, SiV($-$) is the most studied~\cite{Sipahigil2016, Kucsko2013, Hepp2014, SiV_Muller2014, Rogers2014PRL, Jelezko_2014, Jahnke2015a, Neu2013, RiedrichMller:NanoLetters2014, Pingault2014, neu2011, Gali:PRB2013, Goss1996, Dietrich2014}, and stands out with a large Debye-Waller (DW) factor of 0.7, and the demonstration of quantum communication and sensor applications~\cite{Sipahigil2016, Kucsko2013}. On the other hand, the SiV($-$) exhibits short spin coherence times due to phonon dephasing caused by the dynamic Jahn-Teller effect on the orbital doublet~\cite{Jahnke2015a}, thus cooling to the millikelvin regime is required for quantum bit operations~\cite{Becker2018, Sukachev2017}. It is predicted that PbV($-$) might have much longer spin coherence times because of the enlarged gap of the orbital doublet caused by spin-orbit interaction but with the expense of smaller Debye-Waller factor than that of SiV($-$)~\cite{thiering2018emph}. 
	
Alternatively, by removing an electron from $X$V($-$) centers, an orbital singlet with $S=1$ ground state appears~\cite{Iakoubovskii2001, Edmonds2008, Goss1996, Goss2005, Goss2007} that should have intrinsically long coherence times. Recently, it has been demonstrated that SiV($0$) exhibits spin coherence time almost up to a second and relaxation time nearly a minute~\cite{rose2017observation}  at 20~K together with a near-infrared fluorescence signal, and has been proposed for quantum communication applications~\cite{rose2017observation}. This observation naturally shifts the focus towards $X$V($0$) color centers in diamond. However, the nature of the excited and shelving states and levels are far from being understood for SiV($0$)~\cite{rose:arxiv}. In particular, the 946-nm zero-phonon-line (ZPL) optical transition of SiV($0$) (see Refs.~\onlinecite{Allers1995, Breeding2008, DHaenensJohansson2011, rose2017observation, Green2017, green2018electronic}) was originally assigned to an $^3A_{2g} \leftrightarrow {^3A_{1u}}$ electronic excitation from the ground state to the excited state~\cite{DHaenensJohansson2011, Gali:PRB2013}, however it has been very recently revealed that the excited state should be a $^3E_u$ state deduced from stress measurements~\cite{green2018electronic}. Furthermore, a dark $^3A_{2u}$ state below the $^3E_u$ by 6.7~meV was activated in the luminescence spectrum by exerting uniaxial stress on the diamond sample~\cite{green2018electronic}. The optical signals of other $X$V($0$) centers have not yet been identified at all. First principles methods are major tools to explore the complex physics of point defects that can strongly contribute to understanding SiV($0$) color center and identifying the other $X$V($0$) color centers.
	
In this Letter, we present first principles results on the optical properties of $X$V($0$) color centers in diamond. We show that the electrons and phonons are strongly coupled in the electronic excited states, and they constitute of a $(e_g\otimes e_u) \otimes E_g$ product Jahn-Teller (pJT) system, where $e_g$ and $e_u$ refers to the corresponding electronic orbitals, that are simultaneously coupled to quasi-localized $E_g$ symmetry breaking local vibrational mode. This pJT effect is responsible for the anomalous optical spectrum of SiV($0$). We briefly discuss our results in the context of quantum technology applications. We provide the theoretical optical signatures of the other $X$V($0$) color centers too.  
	
$X$V($0$) defect has six carbon dangling bonds and the impurity atoms sits in the inversion center of diamond [see Fig.~\ref{fig:PJT}(c)] and exhibits $D_{3d}$ symmetry. These six dangling bonds introduce $a_{1g}\oplus a_{2u}\oplus e_{u} \oplus e_{g}$ orbitals \cite{Goss1996, Goss2007, DHaenensJohansson2011, Hepp2014, Gali:PRB2013}. The $a_{1g}$ and $a_{2u}$ levels fall in the valence band (VB) of the diamond~\cite{Huler2017}. The $e_{u}$ level is fully occupied by four electrons and it either resonant with the valence band or pops up in the gap by increasing the $X$ atomic number~\cite{thiering2018emph}. The $e_{g}$ level is occupied by two electrons in the band gap of diamond in the ground state. By promoting an electron from the $e_u$ orbital to the $e_g$ orbital (or forming a single hole on both orbitals), the lowest energy optically active and inactive excited states are formed. The 16 electronic configurations from these orbitals are
\begin{equation}
	\label{eq:states} 
	^{2}E_{u}\otimes^{2}E_{g}={}^{3}A_{2u}\oplus{}^{3}A_{1u}\oplus{}^{3}E_{u}\oplus{}^{1}A_{2u}\oplus{}^{1}A_{1u}\oplus{}^{1}E_{u}\text{,}
\end{equation}
where we focus on the 12 dimensional triplet subspace.
	
\begin{figure}[] 
		\includegraphics[width=\columnwidth ]{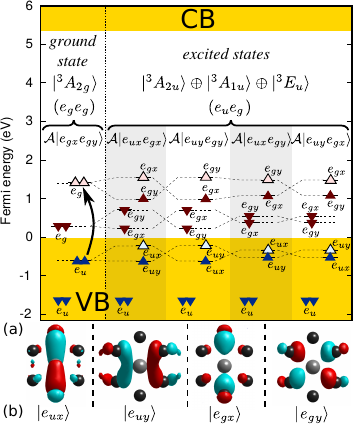} 
		\caption{\label{fig:KS} (a) Kohn-Sham orbitals and levels of the SiV($0$) defect in its $|^3A_{2g}\rangle$ ground state and its four excited state single determinants, as obtained from \emph{ab initio} DFT calculations. The geometry of the system is constrained to $D_{3d}$ symmetry, thus these electronic configurations corresponds to the undistorted $X=0$ configurational coordinate in Fig.~\ref{fig:PJTPES}. The Kohn-Sham orbitals in the spin-up (spin-down) channel are represented by triangles pointing upwards (downwards). The filled (empty) triangles depict occupied (empty) orbitals. The $e_{ux}$, $e_{uy}$ orbitals in the spin-down channel fall into the valence band (VB) and are smeared, so their position is very schematic. However, the $e_{ux}$, $e_{uy}$ orbitals in the spin-up channel form resonant and localized states above the VB edge far from the conduction band (CB). We show the optical excitation path of the $|^3A_{2g}\rangle$ ground state by an inclined arrow pointing upwards. (b) Visualization of single particle Kohn-Sham wavefunctions. The gray ball depicts the impurity atom, whilst the black balls depict the six first neighbor carbon atoms.}
\end{figure}

The excited triplet two-hole wavefunctions and the $|^3A_{2g}\rangle$ ground state can be expressed by the following equations in the hole representation,
\begin{equation}
	\label{eq:WaveTriplets} 
	\!\!\left.\begin{array}{ccc}
	|^{3}A_{1u}\rangle & \!\!=\!\! & \mathcal{A}\frac{1}{\sqrt{2}}\left(|e_{ux}e_{gy}\rangle-|e_{uy}e_{gx}\rangle\right)\!\\
	|^{3}E_{uy}\rangle & \!\!=\!\! & \mathcal{A}\frac{1}{\sqrt{2}}\left(|e_{ux}e_{gy}\rangle+|e_{uy}e_{gx}\rangle\right)\!\\
	|^{3}E_{ux}\rangle & \!\!=\!\! & \mathcal{A}\frac{1}{\sqrt{2}}\left(|e_{ux}e_{gx}\rangle-|e_{uy}e_{gy}\rangle\right)\!\\
	|^{3}A_{2u}\rangle & \!\!=\!\! & \mathcal{A}\frac{1}{\sqrt{2}}\left(|e_{ux}e_{gx}\rangle+|e_{uy}e_{gy}\rangle\right)\!\\
	& \!\!\!\! & \!\\
	|^{3}A_{2g}\rangle & \!\!=\!\! & \mathcal{A}|e_{gx}e_{gy}\rangle\!
	\end{array}\right\} \otimes\left\{ \begin{array}{c}
	\!\!|\!\uparrow\uparrow\rangle\\
	\!\!\mathcal{S}|\!\uparrow\downarrow\rangle\\
	\!\!|\!\downarrow\downarrow\rangle
	\end{array}\right.
	\text{,}
\end{equation}
where we introduce the anti-symmetrization operator $\mathcal{A}|ab\rangle=(|ab\rangle-|ba\rangle)/\sqrt{2} $ and symmetrization operator $\mathcal{S}|ab\rangle=(|ab\rangle+|ba\rangle)/\sqrt{2}$, and spin-up (spin-down) holes are labeled by $\uparrow$ ($\downarrow$). It is worth to note that two singlet levels ($|^1A_1g\rangle$, $|^1E_g\rangle$) for the $e_g^2$ electronic configuration appear above the $|^3A_2g\rangle$ ground state and below the ($e_ge_u$) excited triplet levels of Eq.~\eqref{eq:states}. Additionally, the ($e_ge_u$) excited singlet levels are expected to lie above the ($e_ge_u$) excited triplets due to the Coulomb repulsion~\cite{rose:arxiv, green2018electronic}. We show the single determinant states of the triplets in Eq.~\eqref{eq:WaveTriplets} in Fig.~\ref{fig:KS} in their respective $S_z=+1$ spin substate (see also Supplementary Note 1). We calculate these five $\mathcal{A}|e_{ux}e_{gy}\rangle \otimes |\!\uparrow\uparrow\rangle $ wavefunctions by means of Kohn-Sham density functional theory (DFT) (see Methods).
	
\section*{Results}
\subsection*{Formulation of the product Jahn-Teller\label{sec:PJT} Hamiltonian}
\begin{figure}[] 
		\includegraphics[width=\columnwidth ]{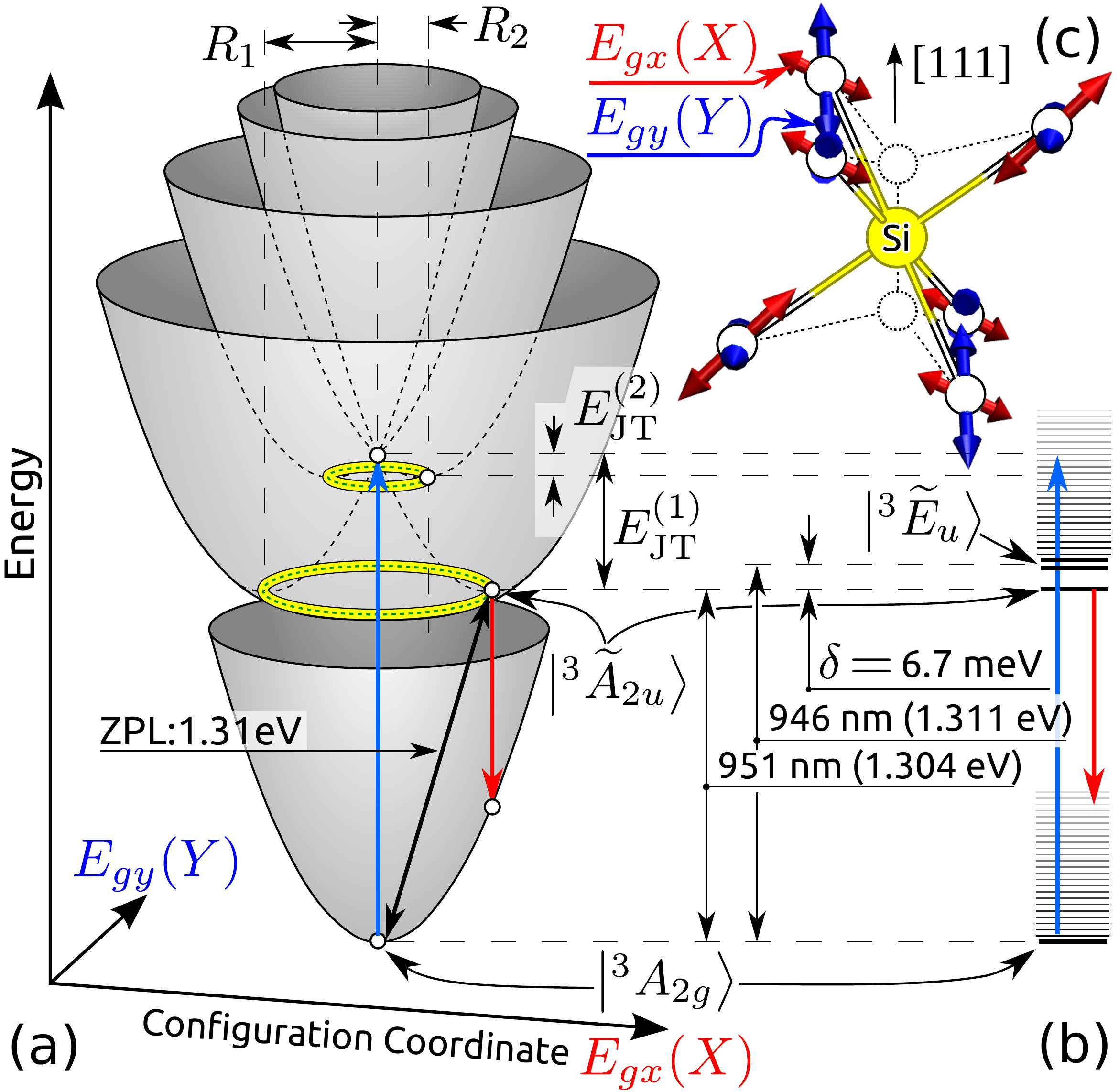} 
		\caption{\label{fig:PJT}(a) Adiabatic potential energy surface of SiV(0) that shows pJT effect in the excited state. We show the vertical absorption and luminescence with blue and red arrows, respectively, whilst the black arrow denotes the ZPL transition. We label the polaronic solutions by tilde in the excited state with energy gap of 6.7~meV. (b) Triplet states of SiV(0). (c) Geometry of the undistorted SiV(0) defect with $D_{3d}$ symmetry. We also depict the $X$ and $Y$ distortion paths of the $E_g$ phonon mode that are quasi-localized on the six carbon atoms.}
\end{figure}

Our DFT calculations indicate a strong Jahn-Teller distortion in the lowest energy triplet excited state of $X$V($0$), going from the high $D_{3d}$ symmetry to the low $C_{2h}$ symmetry. This can be understood by considering the fact that $e_g$ and $e_u$ orbitals are occupied by a single hole in the excited state, thus they are both Jahn-Teller unstable. By applying the $E \otimes e$ Jahn-Teller theory on both orbitals in the \emph{strongly coupled limit} and constructing an antisymmetric product ($\mathcal{A}|e_{ux}^{\varphi}e_{gx}^{\varphi}\rangle$) of the two particles with adding the spin degrees of freedom, one will arrive to the following (see Eqs. S23-S30),
\begin{multline}
	\label{eq:PJTwalk} 
	\bigl|{}^{3}\widetilde{A}_{2u}\bigr\rangle=\mathcal{A}|e_{ux}^{\varphi}e_{gx}^{\varphi}\rangle\otimes\{|\!\uparrow\uparrow\rangle,\mathcal{S}|\!\uparrow\downarrow\rangle,|\!\downarrow\downarrow\rangle\}=\\
	\frac{1}{\sqrt{2}}\left|^{3}A_{2u}\right\rangle 
	-\frac{\cos\left(\varphi\right)}{\sqrt{2}}\left|^{3}E_{ux}\right\rangle
	-\frac{\sin\left(\varphi\right)}{\sqrt{2}}\left|^{3}E_{uy}\right\rangle \text{,}
\end{multline}
which corresponds to walking on the yellow circle of the adiabatic potential energy surface (APES) in Fig.~\ref{fig:PJT}(a). An accurate solution can be found by solving the following Hamiltonian,
\begin{equation}
	\label{eq:H} 
	\hat{H}=\hat{H}_{\mathrm{osc}}+\hat{H}_{\mathrm{pJT}}+\hat{W} ,
\end{equation}
where $\hat{H}_{\mathrm{osc}}$ is the two dimensional harmonic oscillator spectrum of the $E_g$ phonon mode, $\hat{H}_{\mathrm{pJT}}$ is the the pJT Hamiltonian, and $\hat{W}$ is the electron correlation Hamiltonian between the triplet states in Eq.~\eqref{eq:WaveTriplets}. We define $\hat{H}_{\mathrm{osc}}$ by means of ladder operators ($a_{X,Y}$, $a^\dagger_{X,Y}$) as $\hat{H}_{\mathrm{osc}}=\hbar\omega_{E}\sum_{\alpha}^{X,Y}\Bigl(a_{\alpha}^{\dagger}a_{\alpha}+\frac{1}{2}\Bigr)$, where $\hbar\omega_{E}$ is the effective phonon energy.
		
The usual $e \otimes E$ linear Jahn-Teller Hamiltonian~\cite{bersuker2013jahn, bersuker2012vibronic, Ham_1968, Ham:PR1965} is modified to
\begin{multline}
	\label{eq:H_PJT} 
	\hat{H}_{\mathrm{pJT}}=F_{u}\bigl(\hat{X}\hat{\sigma}_{z}\otimes\hat{\sigma}_{0}+\hat{Y}\hat{\sigma}_{x}\otimes\hat{\sigma}_{0}\bigr)\\
	+ F_{g}\bigl(\hat{X}\hat{\sigma}_{0}\otimes\hat{\sigma}_{z}+\hat{Y}\hat{\sigma}_{0}\otimes\hat{\sigma}_{z}\bigr) \text{,}
\end{multline}
where $u/g$ labels the $e_u/e_g$ orbital, and ${\sigma}_{z}$ and ${\sigma}_{x}$ are the standard Pauli matrices in Eq.~\eqref{eq:H_PJT}. $\hat{\sigma}_{0}$ is the two dimensional unit matrix which is introduced for the individual electron-phonon coupling strength $F_u$ and $F_g$~\cite{Qiu2001}.  $(\hat{X},\hat{Y})=(a_{(X,Y)}^{\dagger}+a_{(X,Y)})/\sqrt{2}$ defines the two dimensional configuration space spanned by the $E_g$ vibration mode through the harmonic oscillator ladder operators. See Supplementary Note 3 for details about the derivation of Eq.~\eqref{eq:H_PJT}. Finally, we define the position of electronic levels by $\Lambda$ and $\Xi$ caused by static electronic correlation with the following expression (see Supplementary Note 2),
\begin{multline}
	\label{eq:W} 
	\hat{W}=\Lambda\bigl(|{}^{3}A_{1u}\rangle\langle{}^{3}A_{1u}|-|{}^{3}A_{2u}\rangle\langle{}^{3}A_{2u}|\bigr)\\
	-\Xi\bigl(|{}^{3}E_{ux}\rangle\langle{}^{3}E_{ux}|+|{}^{3}E_{uy}\rangle\langle{}^{3}E_{uy}|\bigr)\text{.}
\end{multline}
	
The overall Hamiltonian of the system in a 4$\times$4 matrix notation is the following by combining Eqs.~(\ref{eq:WaveTriplets}-\ref{eq:W}),
\begin{widetext}
	\begin{equation}
				\label{eq:Hfull}
\hat{H}=\hat{H}_{\mathrm{osc}}+\!\overbrace{\begin{bmatrix}\hat{X}\left(F_{u}+F_{g}\right) & \hat{Y}F_{u} & \hat{Y}F_{g}\\
	\hat{Y}F_{u} & \!\!-\hat{X}\left(F_{u}-F_{g}\right) &  & \hat{Y}F_{g}\\
	\hat{Y}F_{g} &  & \!\!\hat{X}\left(F_{u}-F_{g}\right) & \hat{Y}F_{u}\\
	& \hat{Y}F_{g} & \hat{Y}F_{u} & \!\!-\hat{X}\left(F_{u}+F_{g}\right)
	\end{bmatrix}}^{{\textstyle \!\!\!\!\!\!\!\!\!\!\!\hat{H}_{\mathrm{pJT}}\!:}{\textstyle \mathcal{A}|e_{uy}e_{gy}\rangle\qquad\mathcal{A}|e_{ux}e_{gy}\rangle\qquad\mathcal{A}|e_{uy}e_{gx}\rangle\qquad\mathcal{A}|e_{ux}e_{gx}\rangle}}+\overset{{\textstyle \hat{W_{\,}}}}{\overbrace{\frac{\Lambda}{2}\begin{bmatrix}-1 &  &  & \!\!-1\\
		& 1 & \!\!-1\\
		& \!\!-1 & 1\\
		-1 &  &  & \!\!-1
		\end{bmatrix}-\frac{\Xi}{2}\begin{bmatrix}1 &  &  & \!\!\!-1\\
		& 1 & 1\\
		& 1 & 1\\
		-1 &  &  & 1
		\end{bmatrix}}}
		\text{,}
	\end{equation}
\end{widetext}
where we label the individual single determinant electronic wavefunctions at $\hat{H}_{\mathrm{pJT}}$ for clarity (see also Supplementary Note 3). The diagonal part of the pJT matrix is self-explanatory. If the geometry is distorted towards $+X$, the $|e_{ux}e_{gx}\rangle$ wavefunction would lower its energy, by a joint $F_u+F_g$ coupling strength. In the central part of diagonal Hamiltonian, the two Jahn-Teller effects are destructive, and the joint product Jahn-Teller strength is $F_u-F_g$. 
\begin{figure*}[] 
		\includegraphics[width=\textwidth]{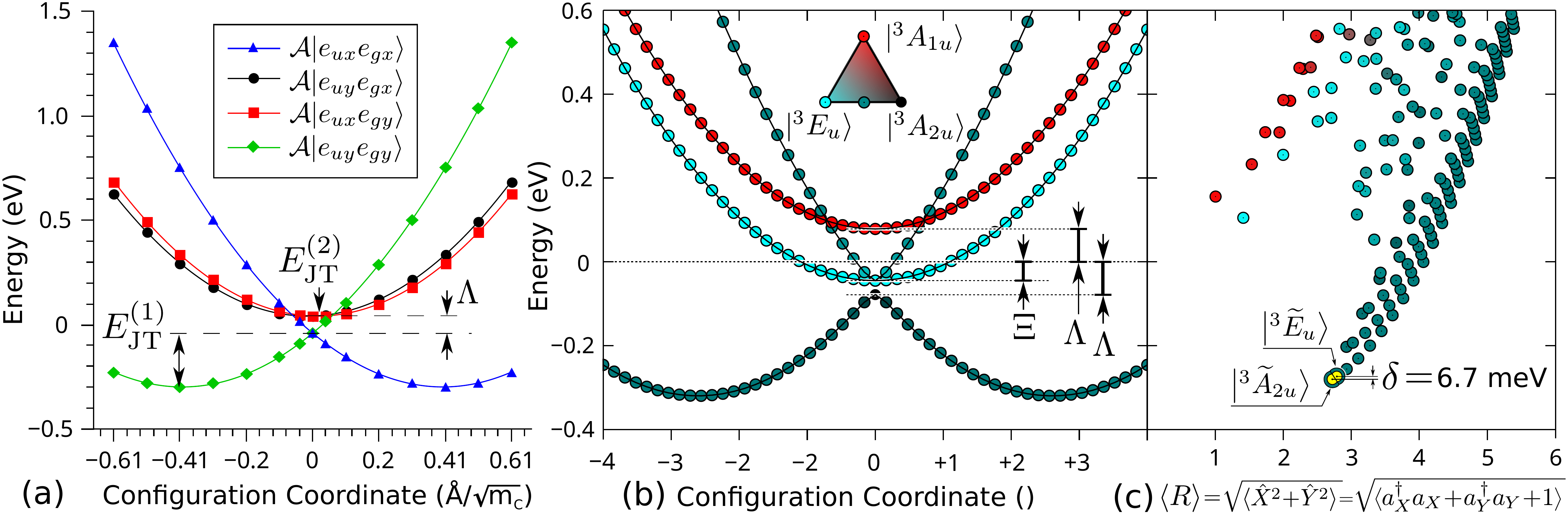} 
		\caption{\label{fig:PJTPES}(a) \emph{Ab initio} APES as obtained from Kohn-Sham DFT for SiV($0$) where $m_\text{c}$ refers to reduced mass of the vibration. The $X=0$ point refers to the geometry relaxation in the constraint of $D_{3d}$ symmetry. The global minimum in APES is obtained upon releasing all symmetry constraints that yields $E_\mathrm{JT}^{(1)}$ JT energy. We mapped the APES with linear interpolation between these two geometries. We mirrored the $X>0$ results to $X<0$ regions. We determined the $E_\mathrm{JT}^{(2)}$ energy by fitting quadratic polynomials on the data points.  (b) Geometry dependence of eigenvalues of Eq.~\eqref{eq:Hfull} by using $X$ as continuous variable and $Y=0$ where the configuration coordinate is in dimensionless unit. We label the wavefunction character with colored balls. Red (teal) balls correspond to the pure $|^3A_{1u}\rangle$ ($|^3E_{ux}\rangle$ or $|^3E_{uy}\rangle$) character that are optically active states. Black balls correspond to the pure $|^3A_{2u}\rangle$ character that is dark state. See Supplementary Note 3 for the four layers in the APES. (c) The convergent polaronic eigenstates of Eq.~\eqref{eq:Hfull} with color coded electronic characters and expectation value of the distortion ($R$) from the $D_{3d}$ symmetry in dimensionless unit. We show the lowest energy vibronic solutions and $\delta$ energy gap between them by yellow balls.}
\end{figure*}

Next, the parameters in  Eq.~\eqref{eq:Hfull} are determined by first principles DFT calculations, and the coupled electron-phonon Hamiltonian is solved (see Supplementary Note 3 for details).

	\subsection*{Parameters from \emph{ab initio} calculations}
	We show the \emph{ab initio} parametrization of the full Hamiltonian for SiV($0$), and discuss the results in detail. The key results are also summarized for the other $X$V($0$) color centers in Table~\ref{tab:ZPLs} as obtained by the same procedure. Our DFT $\Delta$SCF method yields the total energy of the four $\mathcal{A}|e_{uy}e_{gy}\rangle$, $\mathcal{A}|e_{ux}e_{gy}\rangle$, $\mathcal{A}|e_{uy}e_{gx}\rangle$, $\mathcal{A}|e_{ux}e_{gx}\rangle$ electronic configurations (see also Supplementary Note 3 and Fig.~S2). The APES of these states is depicted in Fig.~\ref{fig:PJTPES}(a). This is our starting point to determine the parameters in Eq.~\eqref{eq:Hfull}.

	In $D_{3d}$ symmetry, $E_\text{tot}\bigl[\mathcal{A}|e_{ux}e_{gy}\rangle\bigr]=E_{\text{tot}}\bigl[\mathcal{A}|e_{uy}e_{gx}\rangle\bigr]$ and $E_\text{tot}\bigl[\mathcal{A}|e_{ux}e_{gx}\rangle\bigr]=E_{\text{tot}}\bigl[\mathcal{A}|e_{uy}e_{gy}\rangle\bigr]$, where $E_\text{tot}$ is the DFT total energy (see Supplementary Note 2). Finally, the calculated energy separation is $\Lambda$=78.3~meV for SiV($0$). 
	
	We determine position of $|^{3}E_{u}\rangle$ by a $|e_{u\pm}\rangle=(|e_{ux}\rangle\pm i|e_{uy}\rangle)/\sqrt{2}$ transformation on the defect orbitals, thus two-hole wavefunction will be a single determinant in this basis as $|^{3}E_{u}\rangle=\mathcal{A}|e_{u\pm}e_{u\pm}\rangle$, that can be directly approximated by means of DFT (see Supplementary Note 2). Finally, $\Xi$=44.9~meV is obtained by this procedure for SiV($0$).

	The effective vibration energy $\hbar \omega_E$ can be found by fitting to the parabola of the lowest APES curvature in Fig.~\ref{fig:PJTPES}(a) that results in 75.9~meV for SiV($0$). 
	
	The electron-phonon coupling $F_g$ and $F_u$ parameters can be derived by reading out the characteristic Jahn-Teller energies $E_{\mathrm{JT}}^{(1)}$ and $E_{\mathrm{JT}}^{(2)}$ in the APES [see Fig.~\ref{fig:PJTPES}(a)] as follows,
	\begin{equation}
	\label{eq:EJT}
	E_{\mathrm{JT}}^{(1)}=\frac{(F_{g}+F_{u})^{2}}{2\hbar\omega_E}\text{,}\qquad E_{\mathrm{JT}}^{(2)}=\frac{(F_{g}-F_{u})^{2}}{2\hbar\omega_E}\text{.}
	\end{equation}
	In SiV($0$), $E_{\mathrm{JT}}^{(1)}$=258~meV, that is very significant and seriously affect the calculated ZPL energy. On the other hand, $E_{\mathrm{JT}}^{(2)}$=0.47~meV which is small, and results nearly identical $F_g$ and $F_u$. We note here that we neglect the quadratic Jahn-Teller terms in Eq.~\eqref{eq:H_PJT}, i.e., the APES in Fig.~\ref{fig:PJTPES} is axially symmetric. However, the $\mathcal{A}|e_{uy}e_{gy}\rangle$ has a bit smaller Jahn-Teller energy than that of $\mathcal{A}|e_{ux}e_{gx}\rangle$ by 43, 46, 46, and 48~meV for SiV, GeV, SnV, and PbV, respectively, that would cause a quadratic Jahn-Teller effect. However, these energies are an order of magnitude smaller than that of $E_{\mathrm{JT}}^{(1)}$, and would only lead to minor correction to the results from linear Jahn-Teller approximation. We explicitly proved this for $X$V($-$) color centers in our previous study~\cite{thiering2018emph}. Therefore, we still apply the linear Jahn-Teller approximation for the sake of simplicity.  
	
	Finally, all the parameters could be derived or read out from the calculated APES (see Table~\ref{tab:ZPLs}), thus one can setup the full Hamiltonian in Eq.~\eqref{eq:Hfull}. It is intriguing to use $X$ as a continuous variable at $Y=0$ in Eq.~\eqref{eq:Hfull}, and plot the solution in Fig.~\ref{fig:PJTPES}(b). The contribution of the dark $^3A_{2u}$ state is shown by black balls, while the contribution of the optically active $^3A_{1u}$ state ($z$ polarization) and $^3E_u$ state [$(x,y)$ polarization] is depicted as red and teal balls, respectively. The lowest energy solution will apparently involve the dark $^3A_{2u}$ state. For the full quantum mechanical solution (where $X$ and $Y$ are operators), we use the  following wavefunction ansatz~\cite{Thiering2017SOC, thiering2018emph, thiering2018theory},
\begin{multline}
	\label{eq:Series}
	|\widetilde{\Psi}\rangle=\sum_{n,m}\Bigl[c_{n,m}\mathcal{A}|e_{uy}e_{gy}\rangle+d_{n,m}\mathcal{A}|e_{ux}e_{gy}\rangle+\\
	e_{n,m}\mathcal{A}|e_{uy}e_{gx}\rangle+f_{n,m}\mathcal{A}|e_{ux}e_{gx}\rangle\Bigr]\otimes|n,m\rangle \text{,}
\end{multline}
	where $|n,m\rangle$ is the representation of $E_g$ vibration that we consider up to 15-quanta limit ($n+m \leq 15 $) for the low energy spectrum and up to 50-quanta limit for high energies in Fig.~\ref{fig:PJTPES}(c) as explained in Supplementary Note 3. The vibronic spectrum shows up two deep levels that are separated by $\delta$=6.8~meV, where the deepest level belongs to the vibronic $\bigl|{}^{3}\widetilde{A}_{2u}\bigr\rangle$ and the second level is associated with the vibronic $\bigl|{}^{3}\widetilde{E}_{u}\bigr\rangle$ in SiV($0$).
	
	We also show the derived parameters and results for GeV, SnV, and PbV systems in Table \ref{tab:ZPLs}. There is a clear trend that the ZPL energies increase with heavier impurity atom.
	
	On the other hand the spin-orbit coupling (SOC) will be significant for heavy impurity atoms, and rapidly increase with the atomic number of the impurity atom. Based on our previous calculations for $X$V($-$) defects~\cite{thiering2018emph}, the SOC on $|e_u\rangle$ orbitals,  $\lambda_u$ is 7, 33, 100, and 250~meV for SiV, GeV, SnV, and PbV, respectively. This can be neglected for SiV($0$) but can be significant for the other $X$V($0$) defects, that might alter the ZPL energies SnV(0) and PbV(0). The simultaneous solution for pJT and SOC is out of the scope of this paper but might be required for the ultimately accurate description of the excited states of SnV($0$) and PbV($0$).
\begin{table*}[h] 
	\caption{\label{tab:ZPLs}Calculated parameters of Eq.~\eqref{eq:Hfull} and optical levels of $X$V(0) defects. The $\delta$ is the energy difference between the $\bigl|{}^{3}\widetilde{A}_{2u}\bigr\rangle$ and $\bigl|{}^{3}\widetilde{E}_{u}\bigr\rangle$ states. We note that the ZPL of SnV and PbV may be lowered induced by the spin-orbit coupling that we neglect here. The values inside the parenthesis are experimental data.}
	\begin{ruledtabular}
	\begin{tabular}{lcccc}
		& SiV  & GeV  & SnV  & PbV  \\ \hline 
		$\hbar\omega$ (meV)        & 75.9 & 78.2 & 81.3 & 81.4  \\
		$\Lambda$ (meV)             & 78.3 & 88.6 & 99.5 & 119  \\
		$\Xi$ (meV)                 & 45 & 40 & 42 & 36  \\
		$E_\mathrm{JT}^{(1)}$ (meV)        & 258  & 242  & 217  & 194  \\
		$E_\mathrm{JT}^{(2)}$ (meV)        & 0.47 & 5.18 & 17.2 & 33.4  \\
		$F_g$ (meV)                & 95   & 83   & 67   & 52  \\
		$F_u$ (meV)                & 103  & 112  & 120  & 125  \\
		ZPL$(^3E_u)$ (eV) & 1.34 (1.31\footnote{exp. data from Ref.~\onlinecite{DHaenensJohansson2011}}) & 1.80 & 1.82 & 2.21 \\
		$\delta$ $(^3E_u\leftrightarrow {}^{3}A_{2u})$  (meV)      & 6.7 (6.8\footnote{exp. data from Ref.~\onlinecite{green2018electronic}})  & 7.6 &  9.3  & 10.8 \\
	\end{tabular} 
\end{ruledtabular} 
\end{table*}
	
	\section*{Discussion}
	Experimental data are only available for SiV($0$), thus we can directly compare our results only to them. A recent stress measurement on the photoluminescence (PL) spectrum of SiV($0$) revealed a dark state where the corresponding level was below the ZPL energy by 6.8~meV~\cite{green2018electronic}. Our calculations explain this feature by the pJT effect of the three triplet excited states. The lowest energy branch of the excited state triplets yield $\approx 50 \%$ $|^3 A_{2u}  \rangle$ and  $\approx 50 \%$ $|^3 E_{u}\rangle$ electronic character that can be anticipated from Eq.~\eqref{eq:PJTwalk}. The lowest energy vibronic state is the dark  $\bigl|{}^{3}\widetilde{A}_{2u}\bigr\rangle$ (951~nm), and the next vibronic level above it by 6.7~meV belongs to the optically allowed $\bigl|{}^{3}\widetilde{E}_{u}\bigr\rangle$ (ZPL of 946~nm). We note that in ordinary $e\otimes E$ JT systems, a degenerate $E$ level is the lowest vibronic state quickly followed by a nondegenrate $A$ vibronic level by tunneling splitting energy~\cite{bersuker1963inversion, Reynolds1975, bersuker2013jahn, bersuker2012vibronic, GarcaFernndez2010}. In our present pJT case, the order of these states are reversed, that is a clear signature of the manifestation of the pJT effect.
	
	There are numerous consequences of this finding: (i) The optical polarization of the emitted photons at the ZPL (946~nm) is perpendicular to the symmetry of the axis. (ii) The emission will be strain dependent as symmetry breaking strain can activate the 951-nm ZPL transition, again with photon polarization perpendicular to the symmetry axis of the defect. (iii) The 946-nm ZPL intensity will be temperature dependent, as it depends on the thermal occupation of the $\bigl|{}^{3}\widetilde{E}_{u}\bigr\rangle$ over the lowest energy $\bigl|\widetilde{A}_{2u}\bigr\rangle$. These properties were indeed observed in previous experiments~\cite{DHaenensJohansson2011, green2018electronic}.
	
	We developed a theory for the excited state of $X$V($0$) quantum bits in diamond which revealed a product Jahn-Teller effect, i.e., strong coupling of a localized vibration mode to multiple triplet electronic states. We showed that our theory can explain numerous experimental features of SiV($0$) color center. We predicted the basic optical properties of the other $X$V($0$) color centers too. Our results provide tools to experimentally test product Jahn-Teller systems in solid state. Our findings can be useful to guide experiments on these color centers for quantum bit applications.
	
	\section*{Methods}
	
	\subsection*{Numerical solution of the Jahn-Teller Hamiltonian}
	We determined the energy levels of the product Jahn-Teller system (Eq.~\ref{eq:Hfull}) by a numerical code implemented in GNU octave that we describe in the Supplementary Information.
	
	\subsection*{DFT calculations}
	We characterize $X$V($0$) color centers by plane wave supercell calculations within spin-polarized density functional theory (DFT) as implemented in the \textsc{vasp} 5.4.1 code~\cite{Kresse:PRB1996}. We determine the electronic structure within the Born-Oppenheimer approximation where the ions are treated as classical particles where the minimum energy is found by moving the atoms until the quantum mechanical forces acting on the ions fall below 10$^{-3}$~eV/\AA. We embed the $X$V($0$) defects in a 512-atom diamond supercell. The Brillouin-zone is sampled at the $\Gamma$-point. We applied an energy cutoff at 370~eV for expressing the plane wave basis set within the applied projector-augmentation-wave-method (PAW)~\cite{Blochl:PRB1994, Blochl:PRB2000}. We calculate the excited states with the constrained-occupation DFT method ($\Delta$SCF method)~\cite{Gali:PRB2009}. We used HSE06 hybrid functional~\cite{Heyd03, Krukau06} which reproduces the experimental band gap and the charge transition levels in Group-IV semiconductors within 0.1~eV accuracy \cite{Deak:PRB2010}. For the electron-phonon coupling calculations of $X$V($0$) defects, we apply the same machinery that could well reproduce the ZPL energies of $X$V($-$) exhibiting dynamic Jahn-Teller effect~\cite{thiering2018emph}, thus we expect similar performance for $X$V($0$) too.

	\subsection*{Data availability}
	\makeatletter
	The data from DFT calculations that support the findings of this study are available from Gerg\H{o} Thiering (email: thiering.gergo@wigner.mta.hu) upon reasonable request. \makeatother 
	
	\section*{Acknowledgements}
	Support from \'UNKP-17-3-III New National Excellence Program of the Ministry of Human Capacities of Hungary, the National Research Development and Innovation Office of Hungary within the Quantum Technology National Excellence Program (Project Contract No.\ 2017-1.2.1-NKP-2017-00001), and the European Commission of H2020 ASTERIQS project (Grant No.~820394) is acknowledged. We thank the National Information Infrastructure Development Program for the high-performance computing resources in Hungary. 
	
	\section*{Author Information}
	
	\subsection*{Contributions}
	G.T.\ performed the simulations and analyzed the data. A.G.\ conceived the research and wrote the manuscript with G.T. All authors discussed and commented on the manuscript.
	
	\subsection*{Competing interests}
	The authors declare no competing interests.
	
	\subsection*{Corresponding author}
	Correspondence to Adam Gali.

	\section*{Electronic supplementary material}
	Link to the electronic Supplementary Information.


\end{document}